\documentclass[sigconf]{acmart}

\usepackage{amsfonts}  

\usepackage{booktabs}
\usepackage{hyperref}
 \usepackage{array} 
 \usepackage{fontawesome}
\usepackage{multirow}

\usepackage{colortbl} 
\usepackage{hhline} 
\usepackage{makecell} 
\usepackage{tabularx}
\usepackage{arydshln}
\usepackage{booktabs}
\usepackage{array}
\usepackage{caption} 
\usepackage{lscape}
\usepackage{longtable}
\usepackage{graphicx}
\usepackage{geometry}
\usepackage{lscape}
\usepackage{colortbl}
\usepackage[T1]{fontenc}
\usepackage{tikz}
\usetikzlibrary{positioning}
\usepackage{pifont} 
\usepackage[utf8]{inputenc}
\usepackage{tikz}
\usepackage{pgfplots}
\usepackage{pgfplotstable}
\usepackage{amsmath, amssymb}

\usepgfplotslibrary{polar}
\usepgfplotslibrary{statistics}

\usetikzlibrary{shapes, arrows, positioning, calc}

\usepackage{longtable} 
\usepackage{array} 
\usepackage[normalem]{ulem}
\usepackage{acronym}

\usepackage{multirow}
\usepackage{array}

\usepackage[most]{tcolorbox} 
\usepackage{enumitem}
\usetikzlibrary{shapes.geometric, positioning, arrows.meta, calc}

\tikzset{
    box/.style={
        draw, rounded corners, minimum height=3em, minimum width=10em,
        font=\sffamily\bfseries\large, text centered, fill=#1!30 
    },
    line/.style={draw, thick, -{Triangle[angle=60:1.5mm]}},
    title/.style={font=\sffamily\bfseries\LARGE, text width=3cm, align=center}
}
\definecolor{lightgra}{rgb}{0.4,0.9,0.9}

\newcommand{\yd}[1]{\textcolor{Emerald}{/* #1 (yd) */}}
\newcommand{\yashar}[1]{\textcolor{teal}{[\textbf{Yashar:} #1]}}

\newcommand{\shuai}[1]{\textcolor{gray}{/* #1 (shuai) */}}
\newcommand{\eugene}[1]{\textcolor{darkgray}{/* #1 (eugene) */}}

\renewenvironment{quotation}
{\list{}{\leftmargin=2.5em \rightmargin=0.8em}\item[]}
{\endlist}

\definecolor{bluegreen}{rgb}{0,0.5,0.5}

\newcommand{\dquote}[1]{``#1''}

\AtBeginDocument{%
  }

\copyrightyear{2025} 
\acmYear{2025} 
\setcopyright{cc}
 \setcctype{by}
 \acmConference[SIGIR '25]{Proceedings of the 48th International ACM SIGIR
 Conference on Research and Development in Information Retrieval}{July 13--18,
2025}{Padua, Italy}
 \acmBooktitle{Proceedings of the 48th International ACM SIGIR Conference on
 Research and Development in Information Retrieval (SIGIR '25), July 13--18,
 2025, Padua, Italy}\acmDOI{10.1145/3726302.3730354}
 \acmISBN{979-8-4007-1592-1/2025/07}

\acmISBN{978-1-4503-XXXX-X/2018/06}

\begin{document}

\title[Toward Holistic Evaluation of Generative Recommender System]{Toward Holistic Evaluation of Recommender Systems Powered by Generative Models}



\author{Yashar Deldjoo}
\authornote{Corresponding author.}
\orcid{0000-0002-6767-358X}
\affiliation{%
  \institution{Polytechnic University of Bari}
  \city{Bari}
  \country{Italy}
}
\email{yashar.deldjoo@poliba.it}


\author{Nikhil Mehta}
\orcid{0009-0002-2928-7925}
\affiliation{%
  \institution{Duke University}
  \city{Durham}
  \state{NC}
  \country{USA}
}
\email{nikhil.mehta@alumni.duke.edu}

\author{Maheswaran Sathiamoorthy}
\orcid{0009-0005-2192-3423}
\affiliation{%
  \institution{Bespoke Labs}
  \city{Mountain View}
  \country{USA}}
\email{mahesh@bespokelabs.ai}

\author{Shuai Zhang}
\orcid{0000-0002-7866-4611}
\affiliation{%
  \institution{ETH Zurich}
  \streetaddress{}
  \city{Zurich}
  \country{Switzerland}
}
\email{cheungshuai@outlook.com}

\author{Pablo Castells}
\orcid{0000-0003-0668-6317}
\affiliation{
  \institution{Autónoma University of Madrid}
  \city{Madrid}
  \country{Spain}
}
\email{pablo.castells@uam.es}

\author{Julian McAuley}
\orcid{0000-0003-0955-7588}
\affiliation{%
  \institution{University of California, San Diego}
  \city{La Jolla}
  \country{USA}}
\email{jmcauley@ucsd.edu}

\renewcommand{\shortauthors}{Yashar Deldjoo et al.}


\begin{abstract}
Recommender systems powered by generative models (Gen-RecSys) extend beyond classical item-ranking by producing open-ended content, which simultaneously unlocks richer user experiences and introduces new risks. On one hand, these systems can enhance personalization and appeal through dynamic explanations and multi-turn dialogues. On the other hand, they might venture into unknown territory—hallucinating nonexistent items, amplifying bias, or leaking private information. Traditional accuracy metrics cannot fully capture these challenges, as they fail to measure factual correctness, content safety, or alignment with user intent.

This paper makes two main contributions. First, we categorize the evaluation challenges of Gen-RecSys into two groups: (i) existing concerns that are exacerbated by generative outputs (e.g., bias, privacy) and (ii) entirely \textbf{new risks} (e.g., item hallucinations, contradictory explanations). Second, we propose a holistic evaluation approach that includes scenario-based assessments and multi-metric checks—incorporating relevance, factual grounding, bias detection, and policy compliance. Our goal is to provide a guiding framework so that researchers and practitioners can thoroughly assess Gen-RecSys, ensuring both effective personalization and responsible deployment.

\end{abstract}

\ccsdesc[300]{Information systems~Recommender systems}

\keywords{Generative Recommender Systems, Holistic Evaluation Framework, Hallucination, Factuality, Bias Amplification, Fairness, LLM
}
\maketitle

\section{Introduction}
\label{sec:intro}

For decades, recommender system (RS) evaluation has centered on \emph{held-out} datasets and standardized, non-verbal feedback signals (e.g., clicks, purchases) to measure how well a system retrieves and ranks \emph{known} items \cite{herlocker2004evaluating, canamares2020offline}. In this paradigm, metrics such as NDCG@\(k\), Precision@\(k\), and RMSE typically assume stable user–item distributions and a fixed structured catalog. While effective, this approach inherently treats recommendation as a \emph{surface-level} task of item retrieval and ranking --- a relatively static process with well-defined ground truth in the form of observed user choices.

Large language models (LLMs) have enabled generative recommender systems (Gen-RecSys) to produce more natural and open-ended recommendations, moving beyond simple top-k lists to textual or multi-modal responses that incorporate user instructions and context-specific justifications~\cite{deldjoo2024recommendation}. These systems may directly generate item IDs~\cite{geng2022recommendation,NEURIPS2023_20dcab0f} or leverage a separate retrieval model (e.g., Retrieval-augmented Generation~\cite{deldjoo2024review,fan2024survey}) to retrieve and refine recommendations. However, like many LLM-based applications, Gen-RecSys can be prone to hallucinations, sometimes even recommending out-of-inventory items.  To address challenges such as evolving user needs and query ambiguity,  multiturn conversational Gen-RecSys~\cite{friedman2023leveragingllms} have emerged. By introducing an interactive user-system dialogue, these systems allow for iterative refinement of user intent through clarifications and real-time feedback. This approach transcends static suggestions, offering a more personalized and adaptive recommendation experience where success hinges on retrieval accuracy, maintaining alignment with user intent, and ensuring factual correctness.

A prominent approach is \emph{retrieval-augmented generation (RAG)}~\cite{ram2023iclRAG, LewisPatrick2023RAG, shi2024replug}, a multi-stage approach in which a retriever first collects relevant items, and then an LLM composes the final response. This pipeline can enhance the realism and depth of recommendations, yet it also risks \emph{cascading errors}, since poor retrieval or generation quality in early stages can irreversibly degrade final outputs (such as by introducing hallucinations).

\begin{table*}[!ht]
\caption{Key Differentiators in Evaluation Goals for Traditional vs. Generative Recommender Systems}
\centering
\small
\renewcommand{\arraystretch}{1.}
\begin{tabular}{l| p{5.7cm} p{6.7cm}}
\toprule
\rowcolor{gray!30} \textbf{Aspect} & \textbf{Traditional Recommender Systems} & \textbf{Generative Recommender Systems (Gen-RecSys)} \\
\toprule
\multicolumn{3}{l}{\cellcolor{gray!10}\textbf{I. Overall Objectives \& Output}} \\
\midrule
\textbf{Evaluation Goal} & 
Accuracy-focused (top-$k$ recall, rating prediction) & 
\textbf{Multifaceted}: accuracy plus quality, personalization depth, coherence, and alignment with user context. \\[0.5em]
\textbf{Output Space and Complexity} & 
Structured outputs (e.g., top-$k$ item lists, rating predictions). Mostly static and catalog-bound (against a pre-existing catalog of items), Output text explanation was explored (e.g., with RNN)\cite{zhang2020explainable} & 
Open-ended, multi-modal (text, image, video), and dynamically generated responses. Can create personalized narratives, explanations, and even \textit{hallucinate} items. \\[0.5em]
\midrule
\multicolumn{3}{l}{\cellcolor{gray!10}\textbf{II. Personalization \& Interaction}} \\
\midrule
\textbf{Personalization Granularity} & 
Personalization at an item/rating level, or category level relying on historical collaborative signals; often \textit{coarse-grained}. & 
Enables \textit{aspect-level} personalization (e.g., tone, style, domain context), adapting to individual user instructions, evolving preferences, and persona-level traits. \\[0.5em]
\textbf{Context \& Interaction} & 
Largely \textit{static} scenarios (e.g., single-shot recommendations without multi-turn memory); contextual or temporal aspects often pre-defined and limited to session-based recommendations. & 
Multi-turn, dynamic interactions requiring coherence, contextual updates, and user feedback adaptation over time\footnote{Evaluation must measure the model's ability to maintain consistency over long-term interactions and remember prior user inputs.} \\[1.5em]
\textbf{Evaluating User Satisfaction} & 
Indirect (clicks, ratings). & 
Direct and nuanced assessment (alignment with user's tone, style, factual correctness, conversation quality, intent matching)~\cite{zhang2024personalization} \\[0.5em]
\textbf{Personalization of Tone \& Style} & 
Not relevant or minimal (limited to user preference modeling via IDs or ratings). & 
Highly relevant (LLM must tailor tone, writing style, and content relevance to user's persona and context). \\[0.5em]
\midrule
\multicolumn{3}{l}{\cellcolor{gray!10}\textbf{III. Evaluation Metrics \& Risks}} \\
\midrule
\textbf{Evaluation Metrics} & 
Relevance-oriented metrics (NDCG, Precision) and sometimes novelty/diversity~\cite{castells2021novelty}, as well as explainability~\cite{zhang2020explainable} & 
Traditional metrics + NLP/Gen. metrics (e.g., BLEU, ROUGE, Perplexity, FID, Inception Score, GPTScore~\cite{fu2024gptscore}, and LLM judges~\cite{zheng2024judging}), and persona/alignment metrics. \\[0.5em]
\textbf{Handling Hallucinations} & 
Not applicable (item sets are well-defined; no invented items). & 
Critical (LLM may invent nonexistent items or details), requiring factuality checks and grounding in sources~\cite{tang2024minicheck}. \\[0.5em]
\textbf{Safety and Ethical Concerns} & 
Biases and fairness issues exist but are relatively bounded by catalog constraints; privacy in dataset curation and model training; adversarial robustness and explainability. & 
Bias amplification~\cite{wang2024bias, bao2024decoding}, privacy leaks \cite{kim2024propile}, and new risks (e.g., persona misalignment, hallucinations, unsafe content). Evaluation must incorporate harm detection, bias audits, and scenario-based stress tests. \\[0.5em]
\midrule
\multicolumn{3}{l}{\cellcolor{gray!10}\textbf{IV. Data Dependence}} \\
\midrule
\textbf{Data Reliance} & 
Strongly dependent on historical user-item interactions for personalization. & 
Expands beyond user-item interactions in the training data via knowledge retrieval (e.g., in RAG) and reasoning. \\[0.5em]
\bottomrule
\end{tabular}
\label{tab:eval_comparison}
\end{table*}

\vspace{1mm}
\noindent
\textbf{Traditional vs.\ Generative RS at a Glance.}  
The move from \emph{catalog-bound} systems toward \emph{open-ended} generation brings both new capabilities and new risks. Table~\ref{tab:eval_comparison} compares these paradigms across aspects like output complexity, personalization granularity, and safety concerns. In traditional RS, outputs are often restricted to top-\(k\) item lists, and personalization focuses on item-level preferences learned via user--item interactions. By contrast, Gen-RecSys produce \emph{multifaceted} responses that may incorporate style, tone, and factual justifications, thus requiring more nuanced evaluation metrics. This flexibility also poses serious challenges around factual correctness, potential bias amplification, and handling of private or sensitive information.

Taken together, these emerging Gen-RecSys capabilities introduce an expanded set of \textbf{evaluation challenges}. Conventional accuracy-based metrics cannot capture whether a system’s natural-language justifications are coherent or \emph{factually correct}, nor can they address the risk of generating hallucinations. Ethical implications likewise intensify: what if generative text exposes confidential user details or perpetuates biases? As we dive into these systems, we need to revisit how we define “success” in recommender evaluation, balancing concerns of \emph{helpfulness}, \emph{truthfulness}, and \emph{harmlessness} in an open-ended generative context. This paper brings into focus \textbf{two main contributions}: 
\begin{itemize}[leftmargin=*]
    \item \textbf{A taxonomy of evaluation challenges}, dividing them into \emph{exacerbated} issues (e.g., data bias that intensifies with generative text) and \emph{new} challenges (e.g., item hallucination). 
    \item \textbf{A holistic evaluation framework}, adapted from HELM~\cite{liang2022holistic}, to address the complexities of Gen-RecSys. This includes scenario-based evaluations (e.g., multi-turn dialogues, domain shifts) and multi-metric assessments (e.g., relevance, factuality, fairness). 
\end{itemize}

\noindent \textbf{Reflection on Our Perspective Contribution}
This perspective paper reflects viewpoints of both \textbf{academic} and \textbf{industrial} researchers working at the intersection of recommender systems, LLMs, and generative models, combines classical evaluation with recent advances to propose a novel holistic framework addressing both exacerbated and new risks (e.g., hallucinations, bias, privacy leaks) and defends its stance solely through rigorous literature-based arguments.

\section{Categorizing Challenges in Evaluating Gen-RecSys}
\label{sec:two_group_cat}

We propose a structured framework that organizes the evaluation challenges of generative recommender systems into two categories:
\begin{itemize}
    \item \textbf{Exacerbated Challenges} (cf. Sec.~\ref{subsec:exa_chal}), which intensify existing issues in classical recommenders, and might remain poorly understood in our community.
    \item \textbf{Entirely New Challenges} (cf. Sec.~\ref{subsec:surp}), which are \textbf{entirely new} and introduced by the use of generative models ;
\end{itemize}

Table~\ref{tab:two_group_table} provides an overview of these categories. Below, we discuss each set of challenges in detail.

\subsection{Exacerbated Challenges}
\label{subsec:exa_chal}

\subsubsection*{Accuracy \& Personalization.} Although well-studied, accuracy and personalization become \textit{non-trivial} when the system generates unstructured responses. Traditional offline evaluation (e.g., recall, precision, or NDCG using item IDs) still matters but no longer suffices since the unstructured text generation complicates the notion of \dquote{being correct.} Moreover, large language models (LLMs) must \emph{align} with human preferences or common sense. For instance, a user might refine their travel preference from \emph{\dquote{beach with family activities}} to \emph{\dquote{near a historic town}}, and a generative system should maintain \textbf{groundedness} (factual validity) about specific locations while \emph{aligning} with the new user context. If the textual description of the system is misaligned or references nonexistent landmarks, it fails on both personalization and factual correctness.

\subsubsection*{Novelty, Diversity, \& Coverage.}

In classical recommender systems, integrating diversity can help alleviate filter bubbles \cite{areeb2023filter}.
Generative models can \emph{overtly} highlight novel items or overshadow them, depending on the biases in the training data.
Large Gen-RecSys models can \textit{justify}.
However, measuring whether the system truly achieves \emph{beneficial} novelty can be subjective. Classical measures of catalog-based diversity may not capture how effectively generative text persuades or engages a user to explore new areas.
Techniques such as Diverse Preference Optimization~\cite{lanchantin2025diverse} can be used to improve the diversity of the output of LLMs.

\subsubsection*{Bias Amplification and Emergent Unfairness.}

Gen-RecSys may \emph{exacerbate} popularity bias since they are trained on internet-sourced datasets that inherently reflect real-world popularity distributions~\cite{deldjoo2024understanding,bao2024decoding}. This bias can be further amplified when newer models are trained on synthetic data generated by biased models~\cite{wang2024bias}. Additionally, cultural or demographic stereotypes can appear in textual justifications, and such systems risk perpetuating cultural stereotypes or not respecting cultural norms. For instance, a system trained on diverse internet data might inappropriately recommend alcohol to a user from a culture where it is discouraged. Specialized \emph{fairness and bias} audits must scrutinize not only the recommended items 
but also the \emph{narrative} surrounding them .


\subsubsection*{Privacy \& Security.}
Large language models memorize a lot of information from the Internet and use their memory to answer questions. 
This opens them up for prompting for sensitive questions, such as asking for personally identifiable text \cite{kim2024propile, nakka2024pii}, especially since such information may be present in the training data~\cite{carlini2021extracting}. 
If the information provided by the user is included in the context (e.g., the system prompt) for personalizing information, there is a risk of leakage if adversarial prompts are used. This amplified privacy risk calls for guardrails that detect and redact potential leaks~\cite{li2024llm}; see also~\cite{nazary2025poison} for risks related to text vulnerability, which adversaries could exploit to attack textual (embedding-based) RAG systems.

\subsubsection*{Benchmark Scarcity.}
Classical recommender systems benchmarks (e.g., MovieLens\cite{harper2015movielens}, Amazon product reviews~\cite{he2016ups}) focus on item-level metrics and lack open-ended textual annotations. Evaluating complex generative behaviors (e.g., nuanced explanations) therefore demands new datasets and protocols. While we can still apply item-level accuracy as a baseline, additional dimensions, such as factual correctness, style alignment, or user satisfaction in dialogues, remain largely unaddressed by existing resources.
\vspace{-2.5mm}

\subsubsection*{Efficiency and Cost Considerations.} Generative models are often expensive to train, host, and serve—especially for multi-turn recommendations at scale. A large model may offer higher-quality item suggestions or textual explanations, but it also incurs greater inference times, memory footprints, energy consumption, and monetary costs (e.g., pay-per-token API usage) ~\cite{luccioni2024power, agarwal2023llm, agrawal2024taming}. Evaluations thus must weigh metrics such as throughput, latency, or budget constraints alongside more classical quality measures. Relying on multi-agent designs can exacerbate these costs further, as each agent calls compounds overhead.


\begin{table}[!h]
\centering
\caption{Key Challenges in Two Groups: 
\emph{Exacerbated Challenges} vs.\ \emph{New Challenges}.}
\label{tab:two_group_table}
\small
\begin{tabular}{p{3.7cm} p{4.7cm}}
\toprule
\rowcolor{gray!20}
\textbf{Classical \& Exacerbated Challenges} 
& \textbf{Examples / Notes} \\
\midrule
\textbf{Accuracy \& Personalization} 
& Harder to define “correctness” when outputs are open-ended \\
\textbf{Novelty, Diversity, Coverage} 
& Generative biases overshadow or excessively highlight certain items \\
\textbf{Bias Amplification} 
& Increased risk of demographic/cultural stereotypes in text \\
\textbf{Privacy Risks} 
& LLM may leak private info or infer sensitive attributes \\
\textbf{Benchmark Scarcity} 
& Traditional RS datasets insufficient for free-text generation \\
\textbf{Cost \& Efficiency} 
& Larger LLMs are expensive (token-based or GPU overhead); 
  multi-turn recs compound inference cost \\
\bottomrule
\rowcolor{gray!20}
\textbf{New Challenges} 
& \textbf{Examples / Notes} \\
\midrule
\textbf{Hallucinations / Out-of-Catalog} 
& Model invents items not in the database \\
\textbf{Lack of Reference Data} 
& No single “gold-standard” for open-ended texts or dialogues \\
\textbf{Inconsistent Explanations / Forgetfulness} 
& Multi-turn misalignments; ignoring user constraints mid-conversation \\
\textbf{Persuasion \& User Influence} 
& Generated text can subtly push certain products or viewpoints \\
\textbf{Fine-Tuning \& Continual Updates} 
& Domain norms shift frequently; alignment must be revisited \\
\textbf{Chain-of-Thought Contradictions} 
& Displaying intermediate reasoning can reveal illogical or biased steps \\
\textbf{Multi-Agent Complexity} 
& Multiple LLM modules can cascade errors; 
  agent coordination is non-trivial \\
\bottomrule
\end{tabular}
\end{table}

\subsection{Surprises!}
\label{subsec:surp}

\subsubsection*{Hallucinations \& Out-of-Catalog Items}
Unlike a traditional system that can only recommend known items, a large language model might \emph{hallucinate} options that simply do not exist~\cite{mckenna2023sources, zhang2023language}. For example, it might propose a \dquote{\emph{PalmSky Beach Resort}} that is nowhere in the database. This undermines user trust and makes factual verification an essential component of evaluation.

\subsubsection*{Inconsistent Explanations \& Forgetfulness.}
Generative models risk \emph{contradicting} their own earlier statements or ignoring user preferences over multi-turn interactions. For instance, a user who \dquote{hates horror movies} might later receive a horror recommendation. Tracking cross-turn consistency and memory retention becomes critical for a coherent, reliable user experience.

\subsubsection*{Lack of Reference Data.}
Classic generative metrics (e.g., BLEU \cite{papineni2002bleu}, ROUGE \cite{lin2004rouge}) depend on reference texts. But in recommendation scenarios, especially \emph{personalized} or \emph{open-ended} ones, there may be no single correct explanation or item list. Multiple valid narratives could serve the same user. As a result, human evaluation or LLM-as-a-judge methods (e.g., LLM-based critics with carefully designed rubrics) are often required to evaluate text coherence and appropriateness.

\subsubsection*{Persuasion and User Influence.}
Gen-RecSys can \emph{actively} persuade users --- intentionally or not --- to choose items they might otherwise ignore. This persuasive element goes beyond \dquote{accuracy,} touching on \emph{user well-being} and ethical boundaries. A system might subtly push high-margin products or certain viewpoints. Evaluators thus must assess how the generation influences user decisions and whether it respects user autonomy~\cite{carrasco2024large, altmanpersuation}.

\subsubsection*{Fine-Tuning and Personalization Customization.}
Unlike classical RS trained on fixed item interaction data, Gen-RecSys may require specialized \emph{fine-tuning} (e.g., with Reinforcement Learning from Human Feedback, RLHF~\cite{ouyang2022training}) to align with domain-specific norms, brand guidelines, or user preference profiles. Such customization can yield new challenges in evaluation: each fine-tuned model may behave differently, and continual updates may invalidate previously tested behaviors~\cite{kemker2018measuring}.

\subsubsection*{Chain-of-Thought Contradictions.}
Large language models increasingly rely on multi-step “chain-of-thought” reasoning to handle complex queries. In classical recommenders, intermediate reasoning was never user-facing, 
because we simply did not have such chain-of-thought functionality.
by contrast, some Gen-RecSys now display (or partially reveal) their reasoning to build trust and transparency. 
However, these internal chain-of-thoughts can be contradictory or biased \cite{wei2022chain,lightman2023let}). One argument is that it is acceptable for an internal process to \dquote{hallucinate but recover,} since the user only sees the final \dquote{clean} output.
Regardless, unlike math tasks where intermediate correctness is well-defined, 
recommendations have subjective factors, so verifying each reasoning step is non-trivial. 
Errors in the chain of thought may still yield a superficially correct final recommendation, 
or vice versa. 
When these inconsistencies leak into user-facing explanations, they undermine trust 
and complicate alignment with user preferences. 
This challenge calls for careful pipeline design---e.g., verifying final outputs and/or employing 
partial \dquote{process supervision} to keep the chain-of-thought consistent without overexposing 
all intermediate reasoning to end-users.

\vspace{2mm}
\noindent

\subsubsection*{LLM-as-a-Judge: Pros and Cons in Evaluation}
\label{subsec:llm_judge}
One emerging trend for \emph{evaluating} generative outputs is \textbf{LLM-as-a-Judge}~\cite{zheng2023judging,kim2023prometheus}. Rather than hiring human annotators for each new experiment, a powerful LLM is prompted with an evaluation rubric (e.g., checking for factual correctness) and asked to critique or score generated text. 

However, if the \emph{same} LLM family is used for both generation and judging, the evaluation may suffer from self-reinforcement biases. Furthermore, domain-specific tasks---such as fashion or travel recommendations---often demand specialized knowledge that a general-purpose LLM might lack. Combining LLM-as-a-Judge with external knowledge sources or reference checks can mitigate these pitfalls.

\subsubsection*{Discussion of Multi-Agent LLM-Based Recommenders}
\label{sec:multiagent_discussion}

While \emph{LLM-as-a-Judge} is a method for \emph{evaluation}, a different research thread explores \textbf{multi-agent LLM-based recommenders} as a \emph{system design method}. For instance, one agent retrieves relevant items, another agent generates user-facing explanations, and a third agent cross-checks policy compliance~\cite{park2023generative,liang2022holistic}. Although this approach can modularize tasks, it raises new evaluation complexities:

\begin{itemize}[leftmargin=*]
    \item \textbf{Cascade of Errors:} A retrieval agent’s mistakes can propagate into the generation agent’s text, making debugging difficult.
    \item \textbf{Hidden Internal Messages:} Agents may exchange hidden dialogues or chain-of-thoughts invisible to the user, complicating traceability~\cite{park2023generative}. 
    \item \textbf{Role-Specific Evaluation:} Each agent may require separate metrics for retrieval (NDCG), generation quality (GPTScore), and policy checks (toxicity).
\end{itemize}

\begin{figure*}[!t]
    \centering
    \hspace{-7mm}
\includegraphics[width=0.90\linewidth]{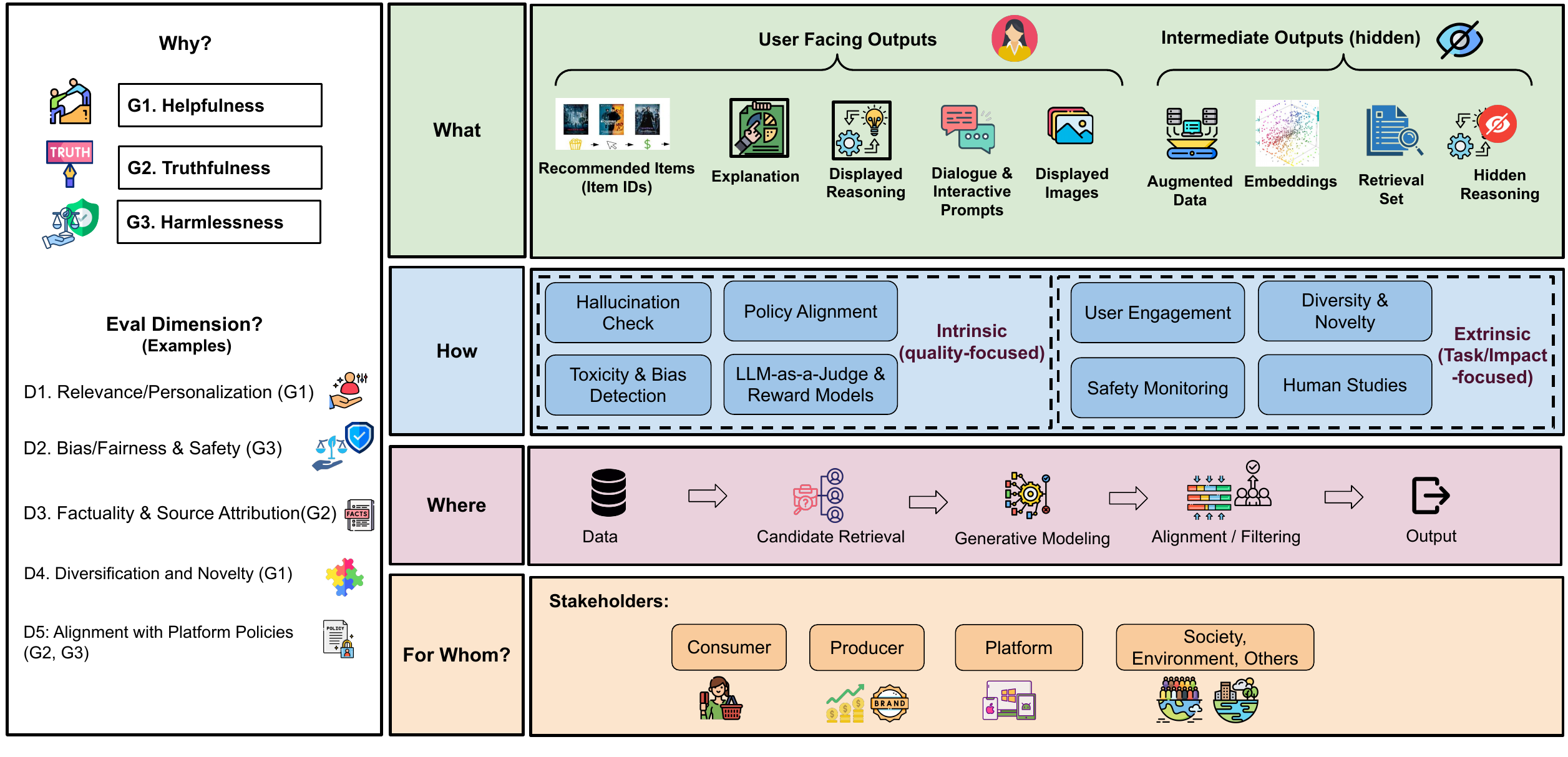}
    \caption{A holistic evaluation framework for Gen-RecSys, linking core goals (helpfulness, truthfulness, and harmlessness) to specific evaluation dimensions (e.g., relevance, fairness, and factuality), along with various factors driven by our proposed WH-question framework.}
    \label{fig:enter-label}
\end{figure*}

\section{Towards Holistic Evaluation of Gen-RecSys}
Classical recommender systems typically output a ranked list of item IDs, often evaluated with metrics such as precision, recall, NDCG, or click-through rates. In contrast, recommender systems powered by generative models \textit{(Gen-RecSys)} can produce \emph{explanations}, \emph{dialogs}, or \emph{multi-modal} content (images, audio). Evaluating these richer outputs demands more than standard item-level metrics. Moreover, Gen-RecSys opens up new risks such as \emph{hallucination}, \emph{biased content}, or \emph{misleading} explanations, requiring a better evaluation paradigm.

\subsubsection*{The Gap in Current Evaluation (Why Classical Metrics Fall Short)}
When systems produce not just recommended items but also text, images, or other user-facing content, they can:
\begin{itemize}
    \item \textbf{Hallucinate} non-existent items or attributes (catalog unawareness),
 \item \textbf{Generate} biased, offensive, or misleading content,
\item \textbf{Lack} a single \dquote{gold-standard} reference for textual explanations or multi-modal data.
Hence, the traditional approach of scoring item IDs based on historical logs (clicks/purchases) no longer suffices. Metrics such as NDCG do not capture the quality or safety of generative output.

\end{itemize}

For Large Language Models, Holistic Evaluation of Language Models (HELM) is a framework for multi-metric comprehensive evaluation \cite{liang2022holistic}. It evaluates models across a broad set of scenarios and metrics so that the performance trade-offs are made clear and evaluation goes beyond a single metric. 
In a similar vein, we advocate a holistic evaluation methodology that addresses both \textbf{extrinsic} (item-ranking) and \textbf{intrinsic} (textual correctness, safety) dimensions. Figure \ref{fig:enter-label} presents a clear visualization of the key evaluation dimensions offered and studied in this work.



\subsection{Defining a WH-Based Evaluation Framework}

Evaluating Gen-RecSys requires a comprehensive approach that incorporates multiple dimensions and perspectives. To guide this evaluation, we propose a structured framework based on six key \dquote{wh-} questions: \emph{What}, \emph{Why}, \emph{How}, \emph{Which}, \emph{Where}, and \emph{For whom}. These questions ensure clarity across various aspects of the evaluation process.

The goal is for each evaluation study to address the following points:  
\emph{(a)} the specific system outputs being evaluated,  
\emph{(b)} the purpose or goal of the evaluation,  
\emph{(c)} the approach used (e.g., human study, automated metrics, LLM-as-a-Judge),  
\emph{(d)} the relevant metrics or dimensions (e.g., relevance, factuality, harmlessness),  
\emph{(e)} the stage or pipeline component being assessed,  
\emph{(f)} the stakeholder group whose perspective is most important.

Table \ref{tbl:evaldim} provides a summary of the key questions, discussed throughout the rest of the paper.

\begin{table*}[h!]
\centering
\caption{WH-Based Evaluation Framework}
\begin{tabular}{@{}ll@{}}
\toprule
\textbf{Question} & \textbf{Explanation} \\ \midrule
\textbf{What} are we evaluating? & The specific outputs of the system (e.g., recommendations, explanations) \\
\textbf{Why} are we evaluating? & The purpose or goal of the evaluation (e.g., accuracy, alignment with user needs) \\
\textbf{How} are we evaluating? & The method of evaluation (e.g., human studies, automated metrics, LLM-based eval.) \\
\textbf{Which metric or dimension} are we evaluating? & The evaluation criteria (e.g., relevance, factuality, fairness) \\
\textbf{Where} in the pipeline are we evaluating? & The specific stage in the recommendation pipeline (e.g., retrieval, generation) \\
\textbf{For whom} is this evaluation intended? & The stakeholders being considered (e.g., end-users, businesses, system designers) \\ \bottomrule
\end{tabular}
\label{tbl:evaldim}
\end{table*}

\subsubsection{(Q1) What to Evaluate? --- Exact Entity}
\label{sec:outputs-definition}

In collaborative filtering (CF), evaluation was primarily focused on retrieval or ranking item IDs against a predefined catalog (cf. Table~\ref{tab:eval_comparison}). Unlike CF, these systems can produce rich outputs. They can:

\begin{itemize}
    \item Generate textual or visual descriptions of items.
    \item Tailor the phrasing of text, considering style and personalization while taking into account factual correctness, groundedness, and safety.
    \item Display additional reasoning steps (\textit{e.g., chain-of-thought explanations}) to improve user trust.
\end{itemize}

Hence Gen-RecSys extends beyond retrieval and ranking and requires redefining the evaluation of the outputs.





As such, the evaluation extends beyond simply measuring which items are recommended—it involves assessing what content is produced and how it supports user decision-making.  Given the diverse types of outputs that a Gen-RecSys can generate, a core challenge in evaluation is \textit{identifying exactly what we are measuring}. We propose a \textbf{structured perspective} by distinguishing between:

\begin{enumerate}
    \item \textbf{User-Facing Outputs} — Content that directly affects the user experience and decision-making.
    \item \textbf{Intermediate Outputs} — Internal representations that influence downstream recommendation tasks but are not directly exposed to the user (e.g., embeddings and reasoning traces).
\end{enumerate}

\subsubsection*{(1) User-Facing Outputs.} User-facing outputs are presented to the users to help in decision making and improving system quality. We categorize Gen-RecSys user-facing outputs as follows:

\begin{enumerate}[label=\textbf{(\alph*)}, leftmargin=3em]
\item \textbf{\boldmath $\hat{R}$: The Recommended Item Set.}
The traditional recommender system output: a ranked list of item IDs. Evaluation uses extrinsic metrics like precision@$k$ and NDCG@$k$, or other relevance-based measures (Section~\ref{subsec:how_evaluate}).

\item \textbf{\boldmath $\hat{E}$: The Explanation.}  
A short textual rationale explaining why these items are recommended. Explanations can improve user trust, highlight key item attributes (e.g., “High in protein,” “Award-winning”), and help users interpret the decision made by the system.

\item \textbf{\boldmath $\hat{Z}$: The Displayed Reasoning Trace.}
An optional snippet of the internal \dquote{chain-of-thought} or reasoning of the model that it chooses to reveal for transparency. In modern models that incorporate \dquote{thinking} processes (e.g., those using chain-of-thought prompting), $\hat{Z}$ illustrates how the model arrived at its recommendation. For example, the DeepSeek-R1 and OpenAI o1 models can be regarded as advanced \dquote{thinking} models (rather than simple autoregressive approaches), demonstrating enhanced problem-solving abilities across diverse tasks by engaging in extended chain-of-thought reasoning.

\end{enumerate}
Reasoning is a nuanced, emerging output that results from the internal workings of large language models (LLMs), often representing a chain of thought or decision-making process behind a recommendation. It differs from an explanation, as reasoning reveals the internal logic of the model and how it arrived at a recommendation, while an explanation simply communicates the rationale in a more user-friendly form. 
\vspace{2mm}

\noindent \textbf{Example -- Fashion Gen-RecSys.}
Suppose a Gen-RecSys provides a set of winter coats, denoted $\hat{R}$, for a user. Along with these coat recommendations, it produces:

\begin{itemize}[leftmargin=1em]
    \item \textbf{Explanation} $(\hat{E})$: "These coats are recommended because they match your navy blue color palette and preference for eco-friendly brands."

    \item \textbf{Reasoning Trace} $(\hat{Z})$:
    Simplified chain-of-thought:
    \begin{enumerate}
        \item User profile indicates preference for navy/dark tones and sustainable brands.
        \item Filtered coat catalog for eco-friendly and navy items.
        \item Ranked by typical price range and brand loyalty.
        \item Selected top three coats aligning with style and sustainability goals.
    \end{enumerate}
\end{itemize}

Fashion Gen-RecSys can recommend diverse items beyond single products, such as complete outfits or home decor, while the core evaluation challenge remains: assessing not only recommended items ($\hat{R}$) but also explanations ($\hat{E}$) and reasoning traces ($\hat{Z}$). \cite{lightman2023let} used process reward models to score individual steps in math reasoning traces. It's an open question of how to do something similar for recommendation reasoning traces.

\subsubsection*{(2) Intermediate (Non-User-Facing) Outputs}
Although the following artifacts are rarely shown to the user, they can be essential for \emph{pipeline-level} or \emph{component-level} evaluation and debugging:

\begin{enumerate}[label=\textbf{(I\arabic*)}, leftmargin=3em]
\item \textbf{\boldmath $\tilde{X}$: Internal User Embedding or Profile Vector.}  
A latent representation capturing user preferences, extracted from interactions or historical data. It controls how recommendations are generated (\emph{e.g.}, clustering users with similar tastes). While invisible to end-users, measuring bias or drift in $\tilde{X}$ is important for fairness and long-term consistency.

\item \textbf{\boldmath $\tilde{C}$: Candidate Retrieval Set.}  
A subset of items retrieved at an earlier stage (e.g., from a vector database or knowledge base) in a RAG-like system. The final recommendations $\hat{R}$ are refined from $\tilde{C}$. Errors or omissions at this stage propagate downstream, so diagnosing \emph{recall, coverage} at the retrieval stage i.e.,  $\tilde{C}$, and \emph{quality} can reveal retrieval bottlenecks.

\end{enumerate}


\subsubsection{(Q2) Why Evaluate? --- Holistic Goals \& Dimensions}
\label{subsec:why_evaluate_holistic}

Evaluating a Gen-RecSys requires balancing several \emph{top-level goals}:
\begin{enumerate}
    \item \textbf{Helpfulness:} Deliver personalized, \emph{useful} recommendations aligned with user context.
    \item \textbf{Truthfulness:} Maintain factual accuracy in items, explanations, or chain-of-thought (avoiding hallucinations or misinformation).
    \item \textbf{Harmlessness:} Prevent harmful, toxic, or biased outputs that could negatively impact users or certain demographics.
\end{enumerate}

To operationalize these goals, we identify \textbf{five} key \emph{dimensions} (D1--D6). Although some dimensions primarily target user-facing text (explanations or chain-of-thought), all can apply to the final recommended items ($\hat{R}$), textual outputs ($\hat{Y}$), or both:

\begin{itemize}
    \item \textbf{D1: Relevance \& Personalization.} 
    Are the recommended items actually relevant to the user's stated preferences (and/or inferred needs), and does the system adapt appropriately to the user context or persona?  Is the system adapting to different user contexts (e.g., cold-start vs.\ power users)?\vspace{0.4mm}
    
   
    \vspace{0.6mm}

    \item \textbf{D2: Bias \& Fairness / Safety.} 
    Do recommendations or generated text disadvantage certain groups (e.g., demographic biases) \cite{deldjoo2024cfairllm}? Are there toxic or hateful elements?


\vspace{0.6mm}
  \item \textbf{D3: Factuality \& Source Attribution.} Are all factual claims in generated text accurate and grounded, and is the knowledge of the system traceable to valid sources (where applicable)? For example, if the system references \dquote{\textit{Destination X has a \$300 average cost,}} is there a credible source to support it, or is it fabricated?
  
  
  \vspace{0.6mm}


    \item \textbf{D4: Diversity \& Novelty.} 
    Does the system move beyond trivial or repetitive suggestions? Are users receiving fresh or serendipitous recommendations?
    

  \vspace{0.6mm}
\item \textbf{D5: Alignment with Platform Policies.} Beyond safe language, does the system obey \textit{domain-specific} or \textit{platform-specific} rules (e.g., age-restricted items, brand and ethical guidelines, or legal constraints)? For example, does it avoid recommending adult-only items or restricted content?

\end{itemize}

Ultimately, these five dimensions let us capture \textbf{why} a generative recommender system should be evaluated, covering user relevance (D1), societal and ethical considerations (D2), factual grounding (D3), content variety (D4), and compliance with policies (D5). Thr techniques for evaluation are discussed in the next section.

\subsubsection{(Q3) How to Evaluate? --- Intrinsic vs.\ Extrinsic Techniques}
\label{subsec:how_evaluate}

Generative RS (Gen-RecSys) produce open-ended text and multi-turn dialogues, making evaluation more challenging than classic item-ranking. 
While human annotation remains a gold standard, it does not scale well~\cite{clark2021all}, 
so recent research has focused on partially or fully automated approaches. 
We group these approaches into \emph{intrinsic} (content-focused) vs.\ \emph{extrinsic} (impact- or task-focused) methods.

\paragraph{Intrinsic Evaluation (Content-Focused).}
Assesses \emph{how} the system communicates:
\begin{itemize}[leftmargin=1em]
    \item \textbf{LLM-as-a-Judge} 
      A strong “judge” LLM can \emph{grade} or \emph{rank} a generative model’s outputs via custom rubrics~\cite{alpaca_eval}.
    \item \textbf{Hallucination Checks.} 
    Compare generated text against a known source to detect invented items or made-up claims (e.g.\ “Hallucination Rate,” knowledge-grounding checks). This is known as grounded factuality.
    \item \textbf{Toxicity \& Bias Detection.} 
    Classifiers like Perspective API, HateXplain, or ToxiGen~\cite{hartvigsen2022toxigen} help spot harmful or stereotyped language. 
    \item \textbf{Policy Alignment.} 
    Automated rule-checkers (e.g.\ GPT-based filters) can catch content that violates brand or legal constraints. 
    
\end{itemize}

\paragraph{Extrinsic Evaluation (Task-Focused).}
Measures whether recommendations \emph{work} for user needs in real scenarios:
\begin{itemize}[leftmargin=1em]
    \item \textbf{User Engagement \& Ranking Metrics.}
    Traditional RS metrics (Precision@k, NDCG@k) or online signals (CTR, dwell time) 
    confirm that \(\hat{R}\) is relevant and useful. 
    \item \textbf{Diversity \& Novelty.}
    Coverage or novelty metrics (e.g.\ Intra-list diversity) can measure whether the system goes beyond safe or obvious picks. 
    \item \textbf{Live Safety Monitoring.}
    Track real user complaints or brand safety violations to ensure the system’s open-ended text remains fair and policy-compliant at scale.
\end{itemize}

\noindent
Overall, robust Gen-RecSys evaluation combines \textbf{intrinsic checks} (to filter toxic, biased, or factually incorrect text) and \textbf{extrinsic metrics} (to ensure meaningful relevance and user satisfaction). As LLMs get stronger, the LLM-as-a-Judge paradigm becomes more promising, but they bring their own challenges as well (e.g., using the same model family for both generations and judging can create self-reinforcing bias). 

\vspace{1em}
\subsubsection{(Q4) Which Metrics or Dimensions?}
\label{sec:wh-which}

In practice, evaluating these final outputs involves both \emph{classical} recommendation metrics and specialized \emph{generative} or \emph{ethical} checks, which can be mapped to our six dimensions (D1--D6). Below are illustrative metric categories:

\begin{itemize}[leftmargin=1em]
    \item \textbf{Classical Metrics (D1, D4).}
    Metrics like NDCG, Recall@k, or coverage remain highly relevant for \(\hat{R}\), helping assess whether the system surfaces user-preferred items (D1: Relevance \& Personalization) and maintains sufficient diversity or catalog exposure (D4: Novelty \& Diversity).

    \item \textbf{Reference-Free Generative Metrics (D3).}
    For textual outputs \(\hat{E}\) or \(\hat{Z}\), methods such as GPTScore can provide \emph{reference-free} evaluations of fluency or style \cite{zhang2024personalization}. More specialized checks—like \emph{hallucination rate}, \emph{factual consistency}, or \emph{source attribution}—align with D3 (Factuality \& Hallucination and/or Groundedness).

    \item \textbf{Multimodality Metrics (D3).}
    When generative outputs include images, videos, or other media (\(\hat{M}\)), metrics like Inception Score (IS) or Fréchet Inception Distance (FID) can gauge \emph{visual} output quality, while user preference experiments can assess relevance. Large-scale platforms (e.g., YouTube) may require synthetic sampling or curated test sets to evaluate such content correctness, and groundednesss (D3)).

    \item \textbf{Ethical \& Societal Benchmarks (D2, D5).}
    To ensure \emph{harmlessness} and \emph{policy alignment}, one can run adversarial tests or use toxicity/bias detection datasets. For instance:
    \begin{itemize}
        \item \textit{Toxicity Classifiers} (D2: Bias \& Fairness / Safety) detect hate-speech or demographic bias in \(\hat{E}\) or \(\hat{Z}\).
        \item \textit{Policy-Adherence Checks} (D5: Alignment with Platform Policies) to confirm no restricted/disallowed items appear in \(\hat{R}\).
    \end{itemize}
\end{itemize}

These categories address \emph{how} we measure each dimension in a user-facing context. Although classical ranking metrics (D1, D4) will always matter for item relevance, the unique challenges of generative recommendations (e.g., hallucinations, toxicity) demand additional methods that capture factual correctness (D3) and robust ethical compliance (D2, D5).

\subsubsection{Where? --- User-Facing Outputs as Our Core Focus}
\label{subsec:where_eval}
Generative recommender systems typically consist of multiple stages, each of which introduces unique evaluation challenges. We propose to assess the system at the following stages:
\begin{itemize}
  \item \textbf{(S1) Data Curation:} The process of collecting and preprocessing user-item data, textual corpora, and external knowledge sources. At this stage, it is important to check for data biases, ensure adequate coverage, and verify that the data meets privacy and policy requirements.
  \item \textbf{(S2) Candidate Retrieval:} The step in which a large pool of items is filtered into a candidate set (often via multi-stage ranking or vector-based retrieval). Here, one must verify that relevant items are captured and that the retrieval does not overemphasize only the most popular items.
  \item \textbf{(S3) Generative Modeling:} The core stage where the language model generates recommendations, explanations, or dialogues. This stage is critical for checking for outcomes, verifying text quality, and 
  ensuring factual correctness.
  \item \textbf{(S4) Alignment / Filtering:} If additional post-training is done with supervised fine-tuning, RLHF, or Reinforcement fine-tuning, the evaluation will be necessary to check for desired outcomes.
  If guardrails (such as policy filters) are applied to refine outputs for compliance with guidelines and safety standards, those need to be evaluated as well.
  \item \textbf{(S5) User-Facing Output \& Feedback:} The final stage, where recommendations and any associated text are presented to users. Here, live user metrics (such as engagement or complaint rates) are essential for confirming that the system meets user expectations and adheres to all quality dimensions.
\end{itemize}

\begin{figure}[!ht]
\centering
\begin{tikzpicture}[font=\footnotesize,>=stealth, node distance=0.3cm]
  \node[draw, rectangle, fill=gray!30, align=center, rounded corners, 
        minimum width=1.3cm, minimum height=0.7cm] (user) {User $u$};

  \node[draw, dashed, thick, rounded corners, fill=blue!15, align=center,
        below=0.6cm of user, text width=3.5cm, minimum height=1.1cm] (nlprofile) 
        {Generate NL Profile (Personalized Text Generation)};

  \draw[->, thick] (user) -- (nlprofile);

  \node[draw, rectangle, align=left, rounded corners, 
        left=0.5cm of nlprofile, text width=2.0cm, minimum height=0.8cm] (smallbox) 
        {\textbullet\ User Attributes\\\textbullet\ User Interactions};

  \draw[->, thick] (smallbox.east) -- (nlprofile.west);

  \node[draw, thick, rectangle, fill=red!20, align=center,
        below=0.8cm of nlprofile, minimum width=0.7cm, minimum height=0.7cm] (xbox) {$X$};

  \draw[->, thick] (nlprofile.south) -- (xbox.north);

  \node[draw, thick, rounded corners, fill=green!20, align=center,
        below=0.8cm of xbox, text width=3.5cm, minimum height=1.3cm] (genrecsys) 
        {Gen-RecSys};

  \draw[->, thick] (xbox.south) -- (genrecsys.north);

  \node[draw, rectangle, fill=purple!20, align=center, 
        minimum width=0.7cm, minimum height=0.7cm, right=1.5cm of genrecsys] (rhat) {$\hat{R}$};

  \draw[->, thick] (genrecsys.east) -- (rhat.west);

  \node[draw, rectangle, fill=yellow!20, align=center, 
        minimum width=0.7cm, minimum height=0.7cm, below=1.5cm of genrecsys] (yhat) {$\hat{Y}$};

  \draw[->, thick] (genrecsys.south) -- (yhat.north);

  \node[align=center, font=\footnotesize, right=1.5cm of yhat, text width=2.0cm] (intrinsicEval) 
  {Extrinsic Eval.\\$E(\hat{Y},Y)$};
  \draw[->, thick] (yhat.east) -- (intrinsicEval.west);

  \node[align=center, font=\footnotesize, above=1.0cm of rhat, text width=2.0cm] (extrinsicEval) 
  {Intrinsic Eval.\\$E(\hat{R},R)$};
  \draw[->, thick] (rhat.north) -- (extrinsicEval.south);

\end{tikzpicture}
\caption{Overview of the Generative Recommender System (Gen-RecSys) pipeline.}
\label{fig:genrecsys_improved}
\end{figure}
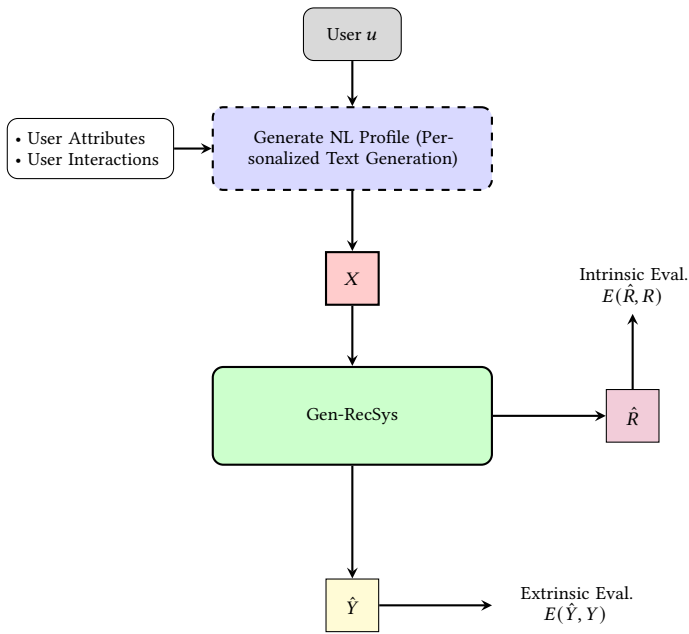

\subsubsection{(Q6) For Whom Are We Evaluating?}
\label{sec:wh-forwhom}
Large-scale recommender systems involve many stakeholders\citep{milano2021ethical, abdollahpouri2021multistakeholder}, and each group cares about specific aspects of evaluation. The stakeholders range from \textbf{end-users} (consumers) to \textbf{producers} (merchants, content creators), \textbf{platforms} (which host the system), and the \textbf{society at large}.
\begin{itemize}[leftmargin=1.2em]
    \item \emph{End-users} value \textbf{relevance} and \textbf{factual correctness}, along with an \textbf{enjoyable} or \textbf{trustworthy} experience.
    \item \emph{Producers} care about \textbf{fair} exposure and brand-safe context. 
    \item \emph{Platform owners} worry about global metrics (engagement, revenue), system scalability, and risk mitigation (e.g., lawsuits if the system is unsafe).
    \item \emph{Societal} concerns include avoiding misinformation, respecting cultural norms, and limiting potential negative externalities (e.g., disinformation or bias).
\end{itemize}

\begin{table*}[!h]
\centering
\small
\setlength{\tabcolsep}{4pt} 
\caption{Scenario Summary: Alignment with Key Dimensions and WH-Questions}
\label{tab:scenario_summary}
\renewcommand{\arraystretch}{0.780}
\begin{tabular}{l|l|l|ccc|l|l|l|l}
\toprule
\textbf{Scenario} 
& \textbf{What?} 
& \textbf{Note} 
& \textbf{Help.} 
& \textbf{Truth.} 
& \textbf{Harm.} 
& \textbf{How?} 
& \textbf{Which?} 
& \textbf{Where?} 
& \textbf{For Whom?}\\
\midrule
\textbf{(A) London} 
& Items only 
& Budget-friendly 
& \cellcolor{green!15}{\checkmark} 
& \cellcolor{red!15}{NA} 
& \cellcolor{green!15}{\checkmark} 
& Extrinsic check 
& Basic ranking 
& Final output 
& End-user \\
\midrule
\textbf{(B) Copenhagen} 
& Items + expl.
& Over-budget; accurate 
& \cellcolor{red!15}{\(\times\)}  
& \cellcolor{green!15}{\checkmark} 
& \cellcolor{green!15}{\checkmark} 
& Extrinsic + intrinsic 
& NDCG, factual  
& Final output 
& User + platform \\
\midrule
\textbf{(C) Berlin} 
& Items + expl. 
& Meets cost; biased text 
& \cellcolor{green!15}{\checkmark}  
& \cellcolor{green!15}{\checkmark}  
& \cellcolor{red!15}{\(\times\)}    
& Intrinsic (tox.) 
& ToxiGen, policy 
& Final output 
& End-user + brand \\
\bottomrule
\end{tabular}
\end{table*}

\subsection{Case Studies of Holistic Evaluation}
\label{sec:holistic_pipeline}

Figure~\ref{fig:genrecsys_improved} illustrates a Gen-RecSys pipeline where a user's attributes and interactions form an NL profile (blue box), yielding an intermediate user representation $X$ (red box) that encapsulates their preferences. The system (green box) then generates a ranked item list $\hat{R}$ and explanatory text $\hat{Y}$ (optionally, $\hat{Z}$ for reasoning). We evaluate $\hat{R}$ extrinsically (e.g., via NDCG) and $\hat{Y}$ intrinsically (e.g., for factuality and bias), ensuring both what is recommended and how it is communicated are appropriately measured.

\begin{table*}[!h]
\small
\centering
\caption{Extended Evaluation Metrics for Gen-RecSys, Mapped to Dimensions (D1--D5)}
\label{tab:extended_metrics}
\footnotesize
\renewcommand{\arraystretch}{0.99}
\begin{tabular}{llll}
\toprule
\textbf{Metric or Technique} 
& \textbf{Purpose / Notes} 
& \textbf{Relevant Dimensions}\\
\midrule

\textbf{RMSE~\cite{herlocker2004evaluating}, MAE~\cite{herlocker2004evaluating}} 
& Prediction accuracy

& D1 (Relevance) \\

\textbf{Recall@k \cite{zhang2023recommender}, Precision@k \cite{zhang2023recommender}, MRR~\cite{voorhees1999trec} } 
& Evaluate candidate retrieval quality 
  in multi-stage pipelines 
& D1 (Relevance)\\&& D4 (Novelty if adapted) \\

\textbf{NDCG@k \cite{jarvelin2002cumulated}, MAP \cite{zhang2023recommender}} 
& Ranking quality and relevance 
& D1 (Relevance)\\ 
&&D4 (Diversity if extended) \\

\textbf{BLEU~\cite{papineni2002bleu}, ROUGE~\cite{lin2004rouge}} 
& Text generation quality (compare to references) 
& D1 (Quality) \\ && Partial D3 (if references are factual) \\

\textbf{GPTScore~\cite{fu2024gptscore}, BARTScore \cite{yuan2021bartscore}} 
& Reference-free scoring of textual outputs  
& D1 (Quality) \\ && can reflect D3 \\

\textbf{Hallucination Rate \cite{ishii2022survey}} 
& Detect nonexistent items or invented facts 
& D3 (Factuality) \\

\textbf{Toxicity Detection (ToxiGen~\cite{hartvigsen2022toxigen}, HateXplain~\cite{mathew2021hatexplain})} 
& Identify harmful or offensive language 
& D2 (Fairness / Safety) \\ && D5 (Policy alignment) \\

\textbf{Bias Audits (Fairness)~\cite{blodgett2020language}} 
& Measure demographic bias, coverage for underrepresented groups 
& D2 (Fairness / Bias) \\

\textbf{Diversity \& Novelty~\cite{castells2021novelty} (e.g.\ ILD, EPD~\cite{castells2011novelty})} 
& Intra-List Diversity, Expected Popularity Complement 
& D4 (Novelty / Diversity) \\

\textbf{Chain-of-Thought Consistency  \cite{wei2022chain, lightman2023let}} 
& Evaluate internal reasoning steps for alignment or contradictions 
& D3 (Factual) \\ &&Possibly D1 or D2 if biases appear \\

\textbf{Live User Metrics (CTR~\cite{joachims2017accurately}, Dwell Time~\cite{lalmas2022measuring})} 
& Extrinsic measure of user engagement and satisfaction 
& D1 (Relevance), + real-time feedback \\
\bottomrule
\end{tabular}
\end{table*}

To show how a single user query can surface various strengths and weaknesses in generative recommendations, 
Consider the following user prompt:
\begin{quotation}
\emph{``Recommend an affordable, eco-friendly travel destination in Europe (under \$500) 
with a good public transport system.''}
\end{quotation}
We examine three hypothetical outputs (London, Copenhagen, and Berlin) that highlight different outcomes in 
\textbf{Helpfulness}, \textbf{Truthfulness}, and \textbf{Harmlessness}. 
Afterward, we provide a summary table capturing the \emph{WH-questions} for each scenario 
and how they align with these evaluation dimensions.

\vspace{1em}
\noindent
\textbf{Scenario (A) --- Item.} \\
\textbf{Item-Only Recommendation}\\
\emph{System Output (Item):} \quad \textit{``London (cost: \$300)''}\\
\emph{Explanation:} \quad (None or minimal)

The recommended item “London” fits the user’s budget, 
and the text claims that it is eco-friendly with free weekend transit. 
From an \emph{extrinsic} perspective (e.g., user cost limit), this is acceptable; 
no hateful or biased phrasing appears.  
\begin{itemize}

  \item {\color{green}\ding{51}} \textbf{Helpfulness:}
    Meets budget, and references eco-friendly aspects.

\item {\color{green}\ding{51}} \textbf{Harmlessness:}
    No toxic or biased language was detected. There is no toxic or biased phrasing to evaluate if no
explanation is given. The recommendations do not point to inappropriate products either.

      \item {\color{gray}\textit{(N/A or Hidden)}} 
    Since no explanatory text is provided, we cannot directly assess the factual correctness of any transit claims related to Truthfulness.
    
\end{itemize}

\vspace{0.5em}
\noindent
\textbf{Scenario (B) --- Item + Explanation.}\\
\emph{System Output (Item):} \quad \textit{``Copenhagen (cost: \$800)''}\\
\emph{System Output (Explanation):} \quad 
\textit{``Copenhagen is renowned for strong sustainability and excellent public transit,
but the total expense exceeds \$500.''}
 
Here, the system states that the cost is \$800, which exceeds the \$500 request 
but accurately reflects Copenhagen’s somewhat higher cost of stay. 
Sustainability and excellent transit are real strengths of the city.  
\begin{itemize}
  \item {\color{red}\bfseries \(\times\) Helpfulness:} 
    Fails the under-\$500 budget constraint.
  \item {\color{green}\ding{51}} \textbf{Truthfulness:}
    Attributes are accurate (\emph{“strong sustainability and excellent transit”}).

    {\color{green}\ding{51}} \textbf{Harmlessness:}
    No offensive, stereotyping, or biased remarks.
\end{itemize}

\vspace{0.5em}
\noindent
\textbf{Scenario (C): Item + Explanation}\\
\emph{System Output (Item):} \quad ``Berlin (cost: \$450)''\\
\emph{System Output (Explanation):} \quad 
\textit{``Berlin meets eco-friendly criteria with a \$40 transit pass. 
Unlike some third-world cities, it has no chaos.''}

The recommendation meets the user’s eco-friendly and cost criteria, 
citing an affordable transit pass. 
However, it references \emph{“third-world cities”} in a negative comparison, 
which introduces a biased or pejorative slant in the explanation text.  
\begin{itemize}
  \item 
      {\color{green}\ding{51}} \textbf{Helpfulness:}
  \item {\color{green}\ding{51}} \textbf{Truthfulness:}
    \$450 total estimate and a “\$40 transit pass” are plausible.
  \item {\color{red}\bfseries \(\times\) Harmlessness:}  
    The phrase \emph{“third-world cities”} is biased, failing D2.
\end{itemize}

Applying our six \emph{WH-questions} reveals that we evaluate both the recommended items ($\hat{R}$) and explanations ($\hat{Y}$) to ensure three key goals: \textbf{helpfulness} (e.g., budget alignment and user context), \textbf{truthfulness} (factual accuracy), and \textbf{harmlessness} (avoiding toxicity or bias). We combine extrinsic checks (such as verifying that the cost does not exceed \$500) with intrinsic analyses (detecting factual errors and harmful language) using classic top-$k$ and cost-based metrics alongside textual evaluations for correctness (D3) and fairness (D2). Our focus is on the final user-facing outputs—though checks may be applied earlier in the pipeline—and the evaluation serves both travelers seeking affordable eco-options and broader platform and ethical responsibilities.

\section{Conclusion and Future Directions}
\label{sec:conclusion}

Generative recommender systems (Gen-RecSys) offer flexible, open-ended outputs that can improve user experiences. Yet they also pose challenges---ranging from hallucinations and privacy leakage to new forms of bias and persuasion. Evaluating such systems demands more than traditional top-$k$ metrics; it requires holistic approaches that measure factuality, alignment, fairness, and safety. In this paper, we presented:
\begin{itemize}[leftmargin=*]
    \item A taxonomy of Gen-RecSys evaluation challenges, distinguishing \emph{exacerbated} and \emph{new} problems.
    \item A multi-dimensional framework, inspired by HELM \cite{liang2022holistic}, that combines classical ranking evaluation with newer generative and ethical checks.
    \item Discussions of emerging methods (multi-agent pipelines, LLM-as-a-Judge) and their associated measurement complexities.
\end{itemize}

\noindent Looking ahead, we see several important directions:
\begin{itemize}[leftmargin=*]
    \item \textbf{Robust Benchmarks:} Building open-source datasets with ground-truth references and annotations (e.g., for factual correctness or bias).
    \item \textbf{Interpretable Guardrails:} Designing policy filters that not only reduce harm but also provide transparent justifications to users.
    \item \textbf{Multi-Stakeholder Perspectives:} Ensuring that producers, platform owners, and end-users all benefit equitably from Gen-RecSys outputs.
\end{itemize}
We believe by can guide Gen-RecSys research to deliver more trustworthy and user-aligned experiences, while mitigating potential harms associated with large language models. Table~\ref{tab:extended_metrics} summarizes extended evaluation metrics for Gen-RecSys by mapping each metric to its purpose and the relevant dimensions (D1--D5).

\bibliographystyle{ACM-Reference-Format}
\bibliography{refs}

\end{document}